\documentclass[twocolumn,showpacs,aps,prl,amsmath,amssymb]{revtex4}

\usepackage{graphicx}
\usepackage{dcolumn}
\usepackage{bm}

\begin{document}

\title{Fractal Weyl laws for chaotic open systems}

\author{W. T. Lu}
\email{wtlu@neu.edu}
\author{S. Sridhar}
\email{srinivas@neu.edu}
\affiliation{Physics Department, Northeastern University,
Boston, MA 02115}

\author{Maciej Zworski}
\email{zworski@math.berkeley.edu}
\affiliation{
Department of Mathematics,
University of California, Berkeley, CA 94720}

\date{\today}

\begin{abstract}
We present a conjecture relating the density of $quantum$ resonances for an
open chaotic system to the fractal dimension of the associated $classical$
repeller. Mathematical arguments justifying this conjecture are discussed.
Numerical evidence based on computation of resonances of systems of $n$
disks on a plane are presented supporting this conjecture. The result
generalizes the Weyl law for the density of states of a closed system to
chaotic open systems.
\end{abstract}

\pacs{05.45.Mt, 03.65.Sq, 05.45.Ac, 31.15.Gy, 95.10.Fh}

\maketitle

The celebrated Weyl law concerning the density of eigenvalues of bound
states is a central result in the spectroscopy of quantum systems \cite%
{Baltes-Hilf}. The Weyl formula states that the asymptotic level number $%
N(k) $, defined as the number of levels with $k_{n}<k$ (where $k\rightarrow
\infty $) is given after smoothing by $N(k)\equiv \#\{k_{n}:\;k_{n}\leq
k\}=Vk^{D}/(D/2)!(4\pi )^{D/2}+...$, for a quantum system bounded in a
region $R$ of $D$-dimensional space whose volume is $V$. For closed systems
with smooth boundaries, the Weyl formula is well-established, and although
primarily valid in the semi-classical limit, nevertheless can be applied
with astonishing accuracy to very low energies extending almost down to the
ground state of integrable and chaotic closed systems. Generalizations of
the Weyl law to other situations have long been sought. A notable example is
the conjecture by Berry \cite{Berry} for the density of states of closed
systems with fractal boundaries, \textit{i.e.} \textquotedblleft fractal
drums\textquotedblright .

Open systems are characterized by resonances defined by complex wavevector $%
\tilde{k}_{n}=\mathrm{Re}(\tilde{k}_{n})+i\mathrm{Im}(\tilde{k}_{n})$,
corresponding to states with finite life times arising from escape to
infinity. Open chaotic systems, which occur in a variety of physical
situations, are generically characterized by a classical phase space
repeller that is fractal. In this letter we present a conjecture relating
the density of resonances for an open chaotic system to the fractal
dimension of the associated \emph{classical} repeller. It can be stated as: 
\begin{equation}
N(k)\equiv \#\,\{\tilde{k}_{n}:\mathrm{Im}\tilde{k}_{n}>-C,\, \mathrm{Re}%
\tilde{k}_{n}\leq k\}\,\,\sim k^{1+d_{H}}  \label{eq:1}
\end{equation}
where $d_{H}$ is the \emph{partial Hausdorff dimension of the repeller} \cite%
[\S 4.4]{Ga}. This relation generalizes the Weyl law for the density of
states of a closed system to chaotic open systems.

In this letter, we will provide a heuristic argument for the validity of
this conjecture, and present new computations that confirm its validity.

The repeller in a scattering problem is defined as the set of points in
phase space which do not escape to infinity at both positive or negative
times. The Hausdorff dimension of the repeller is given by $D_{H}=2d_{H}+2$,
where we did not restrict ourselves to an energy surface. For closed two
dimensional systems, such as compact surfaces of constant negative
curvature, we have real zeros only and $N(k)=\#\{k_{n}:\;k_{n}\leq k\}\sim
k^{2}\,,$ which is consistent with \eqref{eq:1} as $d_{H}=1$ then everything
is trapped.

Our motivation comes from rigorous work on quantum resonances and in
particular from the work of Sj\"{o}strand \cite{Sj} on geometric upper
bounds on their density. The optimal nature of that bound was recently
indicated by a numerical experiment \cite{KL} involving a computation of
quantum resonances for semi-classical Schr\"{o}dinger operators with chaotic
classical dynamics.

Here we consider a different but related problem. Suppose that $Z(k)$ is the
semi-classical Selberg-Ruelle zeta function, with $k$ the wave number. In
some situations the zeros of its meromorphic continuation approximate
semi-classically the quantum resonances of the quantized system -- see \cite%
{Wirzba,WS} for a review of recent theoretical and experimental results.
Because of the exact Selberg trace formula, this is rigorously known for
surfaces of constant negative curvature \cite{PP}. Recently that case was
studied both theoretically and numerically \cite{GLZ} in that setting. It
was shown that a better bound is possible in small energy interval: $\#\,\{%
\tilde{k}_{n}:\mathrm{Im}\tilde{k}_{n}>-C,\,k\leq \mathrm{Re}\tilde{k}%
_{n}\leq k+1\}\,\,\sim k^{d_{H}}$. Bounds such as these normally follow from %
\eqref{eq:1} with a good error estimate. Numerical results for Selberg zeta
functions also indicate that the upper bounds are sharp.

Here we show 
that \eqref{eq:1} holds for zeta functions associated to configurations of
hard disks in the plane. The result is demonstrated by numerical
computations of resonances of $n$-disks on a plane obtained from the poles
of a semi-classical Ruelle zeta function calculated using a cycle expansion.
We have chosen the $n$-disk geometry because it is hyperbolic, has been
treated extensively with well-developed theoretical machinery in the form of
cycle expansions, and has computable fractal dimensions.

The trace formul{\ae } of Selberg, Gutzwiller, and Balian-Bloch (see for
instance \cite{G2} for a survey and references) provide one of the most
elegant and useful ways of expressing the classical quantum correspondence.
To formulate it in terms of the semi-classical zeta function we consider a
semi-classical quantization $\widehat{H}$ of a classical Hamiltonian $H$:
for instance $H=p^{2}+V(q)$ and $\widehat{H}=-\hbar ^{2}\Delta +V(x)$.

The contribution of periodic orbits (PO) to the trace of the resolvent is is
given by 
\begin{equation}
\begin{split}
& \mathrm{tr}\frac{1}{E-\widehat{H}}\big|_{\mathrm{p.o.}}= \\
& \ (\ln Z)^{\prime }(E)(1+\hbar a_{1}(E)+\hbar ^{2}a_{2}(E)+\cdots )\,,
\end{split}
\label{eq:trace}
\end{equation}
where $Z(E)$ is the semi-classical zeta function given by 
\begin{equation}
Z(E)=\exp \left( -\sum_{p}\sum_{n=1}^{\infty }\frac{1}{n}\frac{e^{i\nu
_{p,n}+inS_{p}(E)/\hbar }}{|I-J_{p}^{n}|^{\frac{1}{2}}}\right) \,.
\label{eq:zeta}
\end{equation}
Here $p$'s are primitive periodic orbits, $T_{p}$'s their periods, $\nu
_{p,n}$, the Maslov indices at $n$ iterations, $S_{p}$'s classical actions,
and $J_{p}$'s the linear Poincar\'{e} maps. The formula \eqref{eq:trace} is
somewhat formal but can be made rigorous -- see \cite{SZ} for a recent
presentation in a generalized setting. We also note that, suitably modified,
an exact formula for a class of constant curvature open chaotic systems is
given in \cite{PP}.

For open systems, the quantum resonances are defined as the poles of the
meromorphic continuation of the resolvent (or Green's function) $(E-\widehat{%
H})^{-1}$. Neglecting higher order terms in \eqref{eq:trace} suggests that
in the semi-classical limit the resonances should be approximated by the
complex zeros of the analytic continuation of $Z(E)$. There exists enough
evidence now that this is the case \cite{WS}.

Since a resonance corresponding to $E=E_{0}-i\Gamma $ propagates as $\exp
(-itE/\hbar )=\exp (-itE_{0}/\hbar -t\Gamma /\hbar )$(hence $E_{0}$ is
interpreted as the rest energy, and $\Gamma $ as the rate of decay), the
\textquotedblleft visible\textquotedblright\ resonances should satisfy $%
\Gamma <C\hbar $: if $\Gamma \gg \hbar $ the state decays too fast to be
seen. As a density of resonance states near a given energy level, $E_{0}$,
it is thus natural to consider $N_{\hbar }(E_{0},\delta )=\ \#\{E-i\Gamma
\;:\;|E-E_{0}|<\delta \,,\ \Gamma <C\hbar \}\,.$ From \cite{Sj,KL} we expect
that for hyperbolic classical flows 
\begin{equation}
\begin{gathered} N_{\hbar} ( E_0 , \delta ) \sim h ^{ - d (E_0 , \delta) }
\,, \\ d( E_0 , \delta ) = \frac12 \; {\rm{dim}} \; \{ ( q, p ) \; : \; \\ |
H ( q, p ) -E_0 | < \delta \,, \Phi^t ( q , p ) \not\rightarrow \infty \,, \
t \rightarrow \pm \infty \} \end{gathered}  \label{eq:eff}
\end{equation}%
The set appearing in the definition of $d(E_{0},\delta )$ is the \emph{\
trapped set} or the repeller of the classical flow. It is not clear at this
point what notion of the dimension should be used. The upper bounds \cite{Sj}
and numerical results in \cite{KL} are given in terms of the Minkowski
dimension.

To indicate the reasons behind \eqref{eq:eff} we first recall the standard
argument for obtaining the Weyl law. If $\widehat{H}$ is a quantum
Hamiltonian with a discrete spectrum near $E_{0}$ (which classically
corresponds to the fact that $H^{-1}([E_{0}-\delta ,E_{0}+\delta ])$ is
compact), we have a semi-classical trace formula \cite{DS}, $\mathrm{tr}f(%
\widehat{H})\simeq h^{-D}\int \!\int f(H(p,q))dpdq$. Choosing $f$ close to a
characteristic function of an interval, we obtain the Weyl law. When the set
of closed orbits has measure zero we have a more precise result at
non-critical energies $E_{0}$: $\mathrm{tr}f(\widehat{H}-E_{0})\simeq
h^{-D}\int \!\int f(H(p,q)-E_{0})dpdq$.

Heuristically, the Weyl law for the number of quantum states of a confining
Hamiltonian between energies $a$ and $b$ is obtained by counting the number
of independent quantum states covering the corresponding part of phase
space. By the uncertainty principle, a maximally localized state lives in a
box with sides $\sqrt{h}$ (in phase space) and the number of such boxes
needed to cover the region $a\leq H\leq b$ is proportional to the volume of
the region times $h^{-n}$ where $n$ is the number of the degrees of freedom.
In fact, by using max-min arguments this can easily be made rigorous.

For an open system the volume of the region $a\leq H\leq b$ is infinite but
the density of quantum resonances should be related to the trapped set only,
that is to the set of points which are not carried to infinity by the
classical flow in negative and positive directions. If that set is very
regular in the sense of self similarity, then the number of maximally
localized states needed to cover it is proportional to $h^{-m/2}$ where $m$
is the dimension of the trapped set for energies between $a$ and $b$.
Normally we do not expect the trapped set to be uniformly regular,
especially when we vary the energy, but a reasonable regularity can be
expected from its intersection with a Poincar\'{e} section at a given
energy. Then the covering argument can be applied on the Poincar\'{e}
section, and since the states are invariant under the flow, we obtain the
missing dimension. That gives the infinitesimal density at a given energy in
terms of the dimension of the trapped set at that energy.

For more quantitative variations of this argument (which cannot be made
fully rigorous due to the non-self-adjointness of the problem; only upper
bounds can be obtained following the work of Sj\"{o}strand) see \cite{KL}.
Here we will present an argument based on the trace formula.

For open systems the space in which the trace is taken needs to be modified.
One way to do it is by conjugating $\widehat{H}$ with $\exp (\widehat{G}%
/\hbar )$ where $\widehat{G}$ is a quantization of a Lyapunov function, $%
(d/dt)G\circ \Phi _{t}(q,p)\simeq \rho (q,p)^{2}$, $\rho (q,p)=\text{%
distance to the trapped set}$ \cite{Sj}. For a suitable global choice of $G$%
, the resonances are the eigenvalues of the non-self-adjoint operator $%
\widehat{H}_{G}=e^{-\widehat{G}/\hbar }\widehat{H}e^{\widehat{G}/\hbar }$,
and for $f$'s of the form $\exp (-az^{2}-ibz)$, $a,b>0$, we still have a
relation between the resonances and $\mathrm{tr}f(\widehat{H}_{G})$. We note
that the standard \textquotedblleft Heisenberg picture\textquotedblright\
argument and the property of $G$ show that $\widehat{H}_{G}$ is a
quantization of $H(q,p)-i\rho (q,p)^{2}$. Inserting this, somewhat formally,
to the trace formula above we obtain $\mathrm{tr}f(\widehat{H}_{G})\sim
h^{-D}\int_{H^{-1}(E_{0})}\exp (-\rho (q,p)^{2}/\hbar )dqdp\sim
h^{-D}h^{(2D-1-\mathrm{dim}K_{E_{0}})/2}$, where $K_{E_{0}}$ is the trapped
set in $H^{-1}(E_{0})$, $\mathrm{dim}K_{E_{0}}\simeq 2d(E_{0},\delta )-1$.
This gives the relation \eqref{eq:eff}. We stress that this argument can be
made rigorous as far as the upper bound is concerned.

In view of the semi-classical connection between the zeros of $Z ( E ) $ and
the resonances seen through \eqref{eq:zeta}, it is natural to consider the
analogue of \eqref{eq:eff} for those zeros. It is straightforward but a bit
cumbersome (in view of boundaries) to rewrite the above argument in this
case.

In the case of hard disk scattering in two dimensions (see \cite{WS} and
references given there) the quantum Hamiltonian is given by $-\hbar
^{2}\Delta _{D}$ where $\Delta _{D}$ is the Dirichlet Laplacian. It is then
natural to introduce a new variable $k$, $k^{2}=E/\hbar ^{2}$. Semiclassical
asymptotics correspond to the limit $k\rightarrow \infty $, and the
semiclassical density of resonances considered above should be replaced by $%
N(k)=\#\,\{\tilde{k}_{n}:\mathrm{Im}\tilde{k}_{n}>-C,\,\mathrm{Re}\tilde{k}%
_{n}\leq k\}\,\,.$ The Hausdorff dimension $D_{H}$ of the repeller in %
\eqref{eq:eff} is now independent of the energy level.

For the hard disk geometry, the zeta function can be considered as a
function of $k$ and it takes a somewhat simpler form 
\begin{equation}
\begin{split}
Z(k)& =\exp \left( -\sum_{p}\sum_{n=1}^{\infty }\frac{%
(-1)^{nm_{p}}e^{inkT_{p}}}{n}\frac{1}{|I-J_{p}^{n}|^{\frac{1}{2}}}\right) \\
& =\;\prod_{j=0}^{\infty }\prod_{p}(1-(-1)^{m_{p}}e^{iT_{p}}\Lambda _{p}^{-j-%
\frac{1}{2}})^{j+1}\,,
\end{split}
\label{eq:zetak}
\end{equation}
where $\Lambda _{p}>1$ is the larger of the two eigenvalues of $J_{p}$, and $%
m_{p}$ is the number of reflections of $p$. Here we set particle velocity $%
\upsilon =1$ for simplicity.

Effective ways of evaluating the analytic continuations of semi-classical
(and dynamical) zeta functions have been developed by several authors. The
cycle expansion method \cite{CE} has proved itself to be particularly
succesful. We used it in earlier computations performed for the purpose of
comparisons with experimental data \cite{LS}. As a limit case, one first
considers the two-disk system. In the periodic orbit theory, quantum
resonances form a rectangular lattice in the complex $k$-plane. Thus $%
N(k)\sim k$. This is consistent with $d_{H}=0$ since this system is not
chaotic.

We also chose configurations for which the dimensions of the repellers were
readily available \cite{Ga}: three symmetrically spaced disks of radii $a=1$%
, with centers $r=6$ apart. The zeta function (\ref{eq:zetak}) with smaller $%
j$ will give sharper quantum resonances. For example, for the symmetric
3-disk system with $r=6$, the quantum resonances for the zeta function (\ref%
{eq:zetak}) with $j=0$ are located in the area $\mathrm{Im}k<-0.121$ while
the quantum resonances for the $j=1$ zeta function are located in the area $%
\mathrm{Im}k<-0.699$. We will only consider the zeta-function with $j=0$.
The $C_{3v}$ symmetry of the 3-disk system helps further factorize the zeta
function and makes the periodic orbit theory work more efficiently \cite{CE}%
. There are three irreducible representations, $A_{1}$, $A_{2}$, and $E$.
The cycle expansion with PO up to period 6 gives very accurate calculation
of the quantum resonances of the $A_{1}$ and $A_{2}$ representations for the
range $\mathrm{Im}k>-0.3$ and $0<\mathrm{Re}k<2000$. For the resonances
outside this range, more PO would be needed in the cycle expansion. For
example, including all POs up to period 4, 5, 6, one will get respectively,
26, 33, 39 resonances in the area $2000<\mathrm{Re}k<2010$ and $-0.5<\mathrm{%
Im}k<-0.1$. The quantum resonances of $A_{1}$ and $A_{2}$ representations
are distributed on lines. The $E$ resonances appear to be two-dimensional,
but a careful inspection reveals that they are also distributed on lines.
All these lines tend to move closer to the line $\mathrm{Im}k=-{\frac{1}{2}}%
\gamma _{0}$ as $\mathrm{Re}k$ increases, where $\gamma _{0}$ is the
classical escape rate. This leads to the increase of the density of
resonances \cite{CE}.

\begin{figure}[htbp]
\caption{$A_{2}$ resonances of the 3-disk
system with $r=6$ and $a=1$.}
\label{fig1}
\end{figure}

The principle \eqref{eq:eff} applied to the density $N(k)$ defined before
suggests the law 
\begin{equation}
\log N(k)/\log k\simeq D_{H}/2=1+d_{H}\,,\ \ k\rightarrow \infty \,.
\label{eq:law}
\end{equation}

Representative numerical results are shown in Fig.\ref{fig1} for a 3-disk
configuration with $r=6$ and $a=1$. Fig.\ref{fig2}(a) shows the number of
resonances $N(k)$ vs. $k$ in a strip of width $C=0.28$. Clearly the
dependence of $N(k)$ on $k$ is superlinear. This is confirmed by the
logarithmic plot in the bottom panel which also yields an exponent of $1.288$%
, close to the calculated value of $1+d_{H}=1.2895$. Fig.2(c) shows that the
density of resonances $\Delta N/\Delta C$ peaks at $C\simeq \frac{1}{2}%
\gamma _{0}$, and then decays rapidly for large $C$. This demonstrates that
the range of $C$ used in the present study is more than adequate.

Similar calculations of the quantum resonances with different separation $%
r=5,8,10,12$ were also carried out for the 3-disk system. They all confirm
the validity of \eqref{eq:law} for the symmetric three-disk configurations.

\begin{figure}[htbp]
\caption{(Top) The counting function, $N(k)$, for width $C=0.28$ 
for the resonances in Fig.1. (Middle) The plot of 
$\log N(k)$ against $\log k$. The least square approximation slope is equal
to $1.288$. (Bottom) Dependence of density of resonances $\Delta N/\Delta C$
on strip width $C$. The vertical line is $\frac{1}{2}\protect\gamma _{0}$.}
\label{fig2}
\end{figure}

To be valid the result of this paper requires that the exponent of $k$
should be independent of the counting strip $C$. The numerical calculations
indeed confirm this as shown in Fig.\ref{fig3}, where only weak dependence
of the exponent on the choice of $C$ is observed as $C$ is increased.

We have further checked the result by varying the disk separation. We also
see that the agreement with the dimension persists when we change the disk
separation (Fig.3). To combine data for different $r/a$ values in a single
plot, we rescaled the strip width by $\frac{1}{2}\gamma _{0}$. This is
motivated by the fact that the density of resonances peaks at $C\simeq \frac{%
1}{2}\gamma _{0}$.

\begin{figure}[htbp]
\caption{Dependence of exponent on the
rescaled strip width, $2C/\protect\gamma _{0}$, for the 3-disk system in
three cases with $r/a=5,\,6,10$. $\protect\gamma _{0}=0.4703,0.4103,0.2802$
is the corresponding classical escape rate. The solid lines are the
corresponding Hausdorff dimensions $d_{H}$=0.3189, 0.2895, 0.2330.}
\label{fig3}
\end{figure}

The computational results presented here produce strong evidence for the
connection between the density of resonant states and the fractal dimension
of the repeller on which they concentrate. It is quite possible that
different dimensions might occur in lower and upper bounds once less
symmetric examples are considered. Since zeta functions were used in the
computation, we have also provided evidence that the density of zeros of
zeta functions is related to the dimension of the repeller.

It is worth noting that while fractality arises from $R$ or its boundary $%
\partial R$ in the works of Ref.\cite{Berry}, in the present work both $R$
and $\partial R$ are smooth and instead the \emph{classical phase space is
fractal}.

The connections described here between quantum spectral properties and
classical fractal properties of the associated repeller of open chaotic
systems parallels a similar connection established earlier between the
quantum spectral autocorrelation and the classical decay rate \cite{LS,WS}.
The spectral autocorrelation of the quantum microwave spectra of $n$-disk
billiards was shown to display the fingerprints of the \emph{classical
Ruelle-Pollicott (RP) resonances} $\tilde{\gamma}_{n}=\gamma _{n}\pm i\gamma
_{n}^{\prime }$. The leading RP resonance $\gamma _{0}$ ($\gamma
_{0}^{\prime }=0$) represents the classical decay rate and is related to the
information dimension $d_{I}$ of the repeller by $\gamma _{0}=\lambda
(1-d_{I})$, where $\lambda $ is the average Lyapunov exponent. The higher
RP\ resonances $\tilde{\gamma}_{n}$ with $n>0$ represent fine structure
properties of the fractal repeller comprised of the manifold of trapped
orbits.

In conclusion, the present work establishes a fundamental connection between
a quantum spectral property, and the fractal phase space structure of the
corresponding classical dynamics. The result has important implications for
a variety of areas where it is important to know the number of degrees of
freedom of quantum chaotic systems, and since open quantum chaotic systems
are relevant to a wide range of problems in atomic and chemical systems \cite%
{Delos,Wirzba,G2} and quantum computation.

Work by WTL and SS supported by NSF-PHY-0098801. Work by MZ supported by
NSF-DMS-0200732.


\begin{thebibliography}{99}
\bibitem{Baltes-Hilf} H.P.Baltes and E.R.Hilf, \emph{Spectra of Finite
Systems}, B-I Wissenschaftsverlag, Mannheim (1978).

\bibitem{Berry} M.V.Berry in \emph{Structural Stability in Physics}, eds W. G%
\"{u}ttinger and H. Eikemeier, Springer, p.51-3.

\bibitem{Ga} P. Gaspard, \emph{Chaos, scattering and statistical mechanics,}
Cambridge Univ. Press, 1998.

\bibitem{Sj} J. Sj\"{o}strand, Duke Math. J. \textbf{60} (1990), 1; M.
Zworski, Inv. Math. \textbf{136}, 353 (1999); Notices Amer. Math. Soc. 
\textbf{46}, 319 (1999), J. Sj\"{o}strand, in \emph{Microlocal analysis and
spectral theory} (Lucca, 1996), 377--437, NATO Adv. Sci. Inst. Ser. C Math.
Phys. Sci., 490, Kluwer Acad. Publ., Dordrecht, 1997.

\bibitem{KL} K. Lin, J. Comp. Phys. \textbf{176} (2002), 295; K. Lin and M.
Zworski, Chem. Phys. Lett. \textbf{355}, 201 (2002).

\bibitem{Wirzba} A.Wirzba, Physics Reports, \textbf{309}, 1-116 (Feb. 1999) .

\bibitem{WS} W.T. Lu, K. Pance, P. Pradhan, and S. Sridhar, Physica Scripta 
\textbf{T90}, 238 (2001).

\bibitem{PP} S.J. Patterson and P. Perry, Duke Math. J. \textbf{106}, 321
(2000).

\bibitem{GLZ} L. Guillop\'{e}, K.Lin, and M. Zworski, to appear. \texttt{%
http://xxx.lanl.gov/list/math.MP/new}.

\bibitem{G2} P. Gaspard, D. Alonso, and I. Burghardt, in \emph{Advances in
Chemical Physics XC,} I. Prigogine and S.A. Rice, eds. John Wiley \& Sons,
1995.

\bibitem{SZ} J. Sj\"{o}strand and M. Zworski, J. Math. Pure Appl. \textbf{81}%
, 1 (2002).

\bibitem{DS} M. Dimassi and J. Sj\"{o}strand, \emph{Spectral asymptotics in
the semi-classical limit,} Cambridge Univ. Press, 1999.

\bibitem{CE} P. Cvitanovi\'{c} and B. Eckhardt, Phys. Rev. Lett. \textbf{63}
(1989), 823; P. Cvitanovic and B. Eckhardt, Nonlinearity \textbf{6}, 277
(1993); B. Eckhardt and G. Russberg, Phys. Rev. E \textbf{47}, 1578 (1993).

\bibitem{LS} K.Pance, W.T.Lu and S.Sridhar, Phys. Rev. Lett. \textbf{85},
2737 (2000).

\bibitem{Delos} M. R. Haggerty, J. B. Delos, Phys. Rev. A \textbf{61},
053406 (2000).
\end{thebibliography}
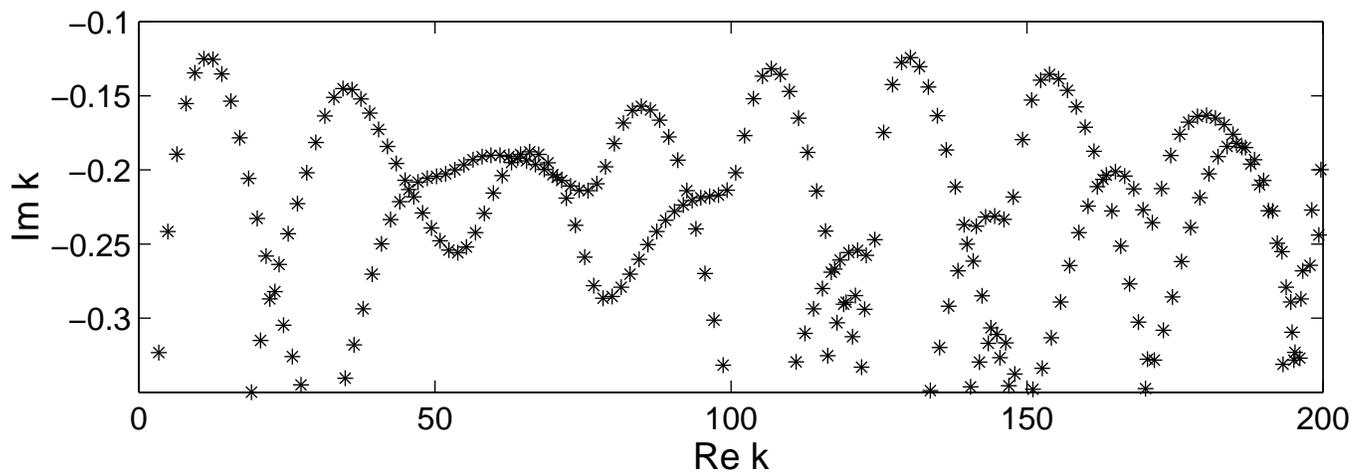
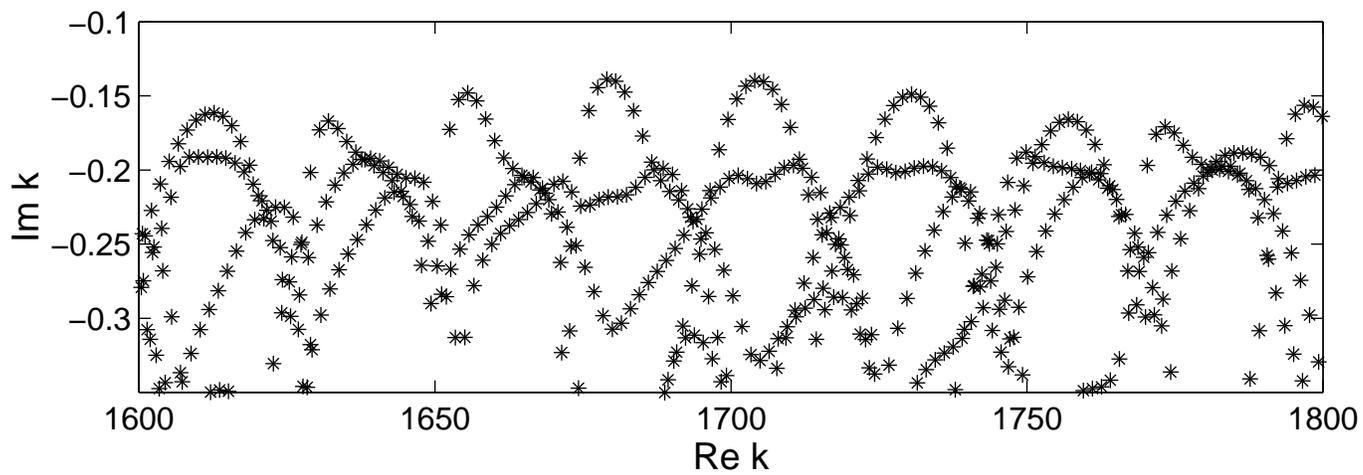
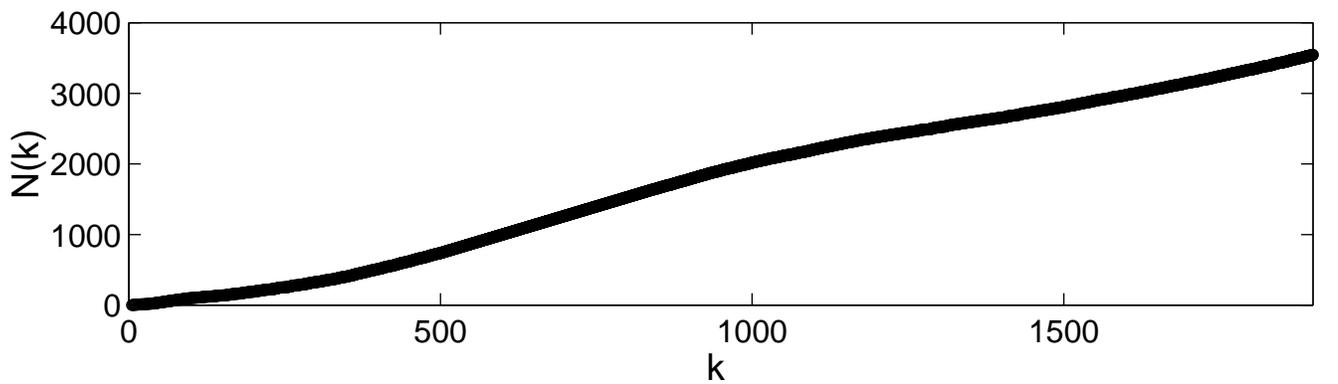
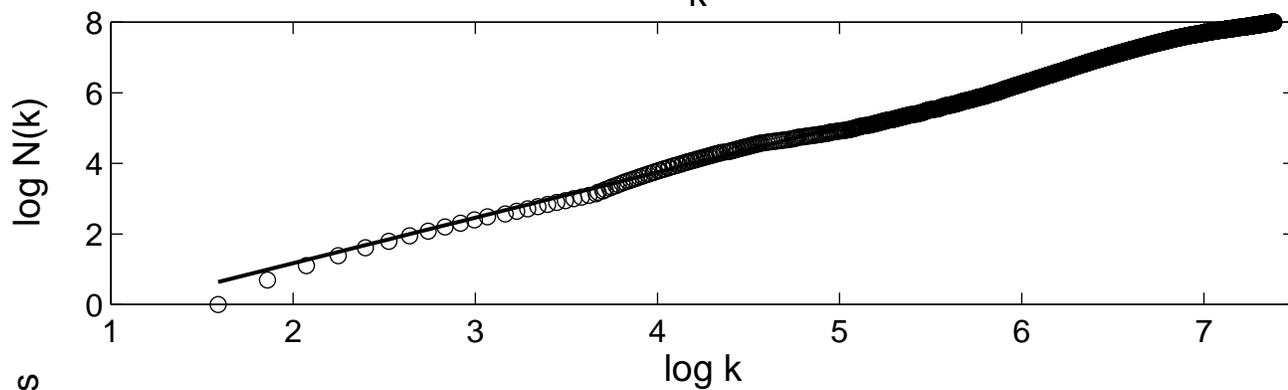
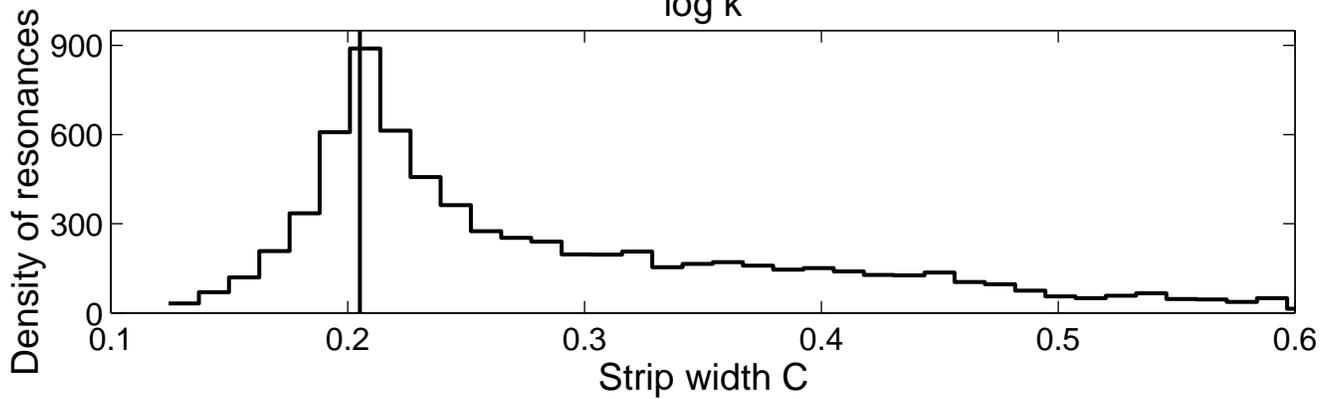
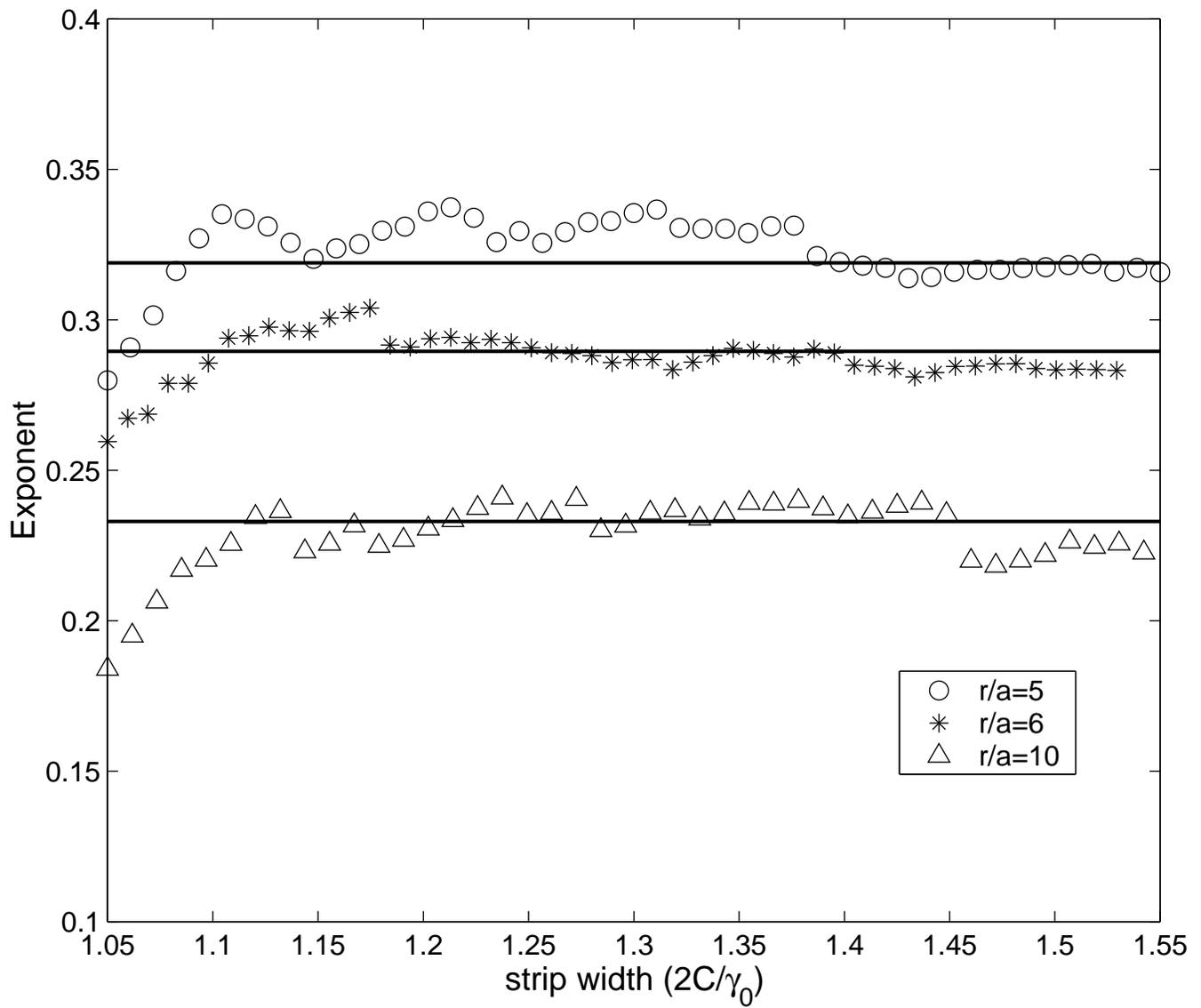
\end{document}